\documentclass{elsart}
\usepackage{natbib,epsfig, amssymb}
\newcommand{\ie}{{\it i.e.}}
\newcommand{\etc}{{\it etc...}}
\newcommand{\eg}{{\it e.g.}}
\newcommand{\cpp}{C${}^{++}$}
\newcommand{\BaBar}{Ba$\overline{{\rm B}}$ar}

\begin{document}
\runauthor{Cranmer, Kyle}
\begin{frontmatter}
\title{Kernel Estimation in High-Energy Physics}
\author[Cranmer]{Kyle Cranmer\thanksref{NSF}}
\thanks[NSF]{This work was partially supported by a graduate research fellowship from the National Science Foundation and US Department of Energy grant DE-FG0295-ER40896}

\address[Cranmer]{University of Wisconsin-Madison}
\begin{abstract}
  Kernel Estimation provides an unbinned and non-parametric estimate
  of the probability density function from which a set of data is
  drawn.  In the first section, after a brief discussion on parametric
  and non-parametric methods, the theory of Kernel Estimation is
  developed for univariate and multivariate settings.  The second
  section discusses some of the applications of Kernel Estimation to
  high-energy physics.  The third section provides an overview of the
  available univariate and multivariate packages.  This paper
  concludes with a discussion of the inherent advantages of kernel
  estimation techniques and systematic errors associated with the
  estimation of parent distributions.

\end{abstract}
\end{frontmatter}

\section{Introduction}

\subsection{Overview}\label{SS:Overview}

Perhaps the most common practical duty of a particle physicist is to
analyze various distributions from a set of data $\{t_i\}$.  The typical tool
used in this analysis is the histogram.  The role of the histogram is
to serve as an approximation of the parent distribution, or
probability density function (pdf) from which the data were drawn.
While histograms are straightforward and computationally efficient,
there are many more sophisticated techniques which have been developed
in the last century.  One such method, kernel estimation, grew out of
a simple generalization of the histogram and has proved to be
particularly well-suited for particle physics.

In order to produce continuous estimates $\hat{f}(x)$ of the parent
distribution from the {\it empirical probability density function} ${\rm epdf}(x) = \sum_i \delta(x-t_i)$, several techniques have been developed.  These
techniques can be roughly classified as either parametric or
non-parametric.  Essentially, a parametric method assumes a model
$\hat{f}(x; \vec{\alpha})$ dependent on the parameters $\vec{\alpha} =
(\alpha_1, \alpha_2, \alpha_3, \dots)$.  The specification of this
model is ``entirely a matter for the practical [physicist]%
\footnote{From a debate between R.A. Fisher and Karl Pearson}''.  The
goal of a parametric estimate is to optimize the parameters $\alpha_i$
with respect to some goodness-of-fit criterion (\eg\ $\chi^2$,
log-Likelihood, \etc).  Parametric models are powerful because they
allow us to infuse our model with our knowledge of physics.  While
parametric methods are very powerful, they are highly dependent on the
specification of the model.  Parametric methods are clearly not
practical for estimating the distributions from a wide variety of
physical phenomena.

The goal of non-parametric methods is to remove the model-dependence of
the estimator.  Non-parametric estimates are concerned directly with
optimizing the estimate $\hat{f}(x)$.  
The prototypical non-parametric density estimate is the histogram%
\footnote{The name `histogram' was coined by Karl Pearson}.  Somewhat
counterintuitively, non-parametric methods typically involve a large -
possibly infinite - number of ``parameters'' (better thought of as
degrees of freedom).  Scott and Terrell supplied a more concrete
definition of a non-parametric estimator, ``Roughly speaking,
non-parametric estimators are asymptotically local, while parametric
estimators are not.''\cite{Scott} That is to say, the influence of a
data point $t_i$ on the density at $x$ should vanish asymptotically
(in the limit of an infinite ammount of data) for any $| x -t_i | > 0$
in a non-parametric estimate.  The purpose of this paper is to
introduce the notion of a kernel estimator and the inherent advantages
it offers over other parametric and non-parametric estimators.

\subsection{Kernel Estimation}
The notion of a kernel estimator grew out of the asymptotic limit of
Averaged Shifted Histograms (ASH).  The ASH is a simple device that
reduces the binning effects of traditional histograms.  The ASH
algorithm is as follows: First, create a family of $N$
histograms, $\{H_i\}$, with bin-width $h$, such that the first bin of
the $i^{\rm th}$ histogram is placed at $x_0 + i h/N$.  Because $x_0$
is an artificial parameter, each of the $H_i$ is an equally good
approximation of the parent distribution.  Thus, an obvious estimate
of the parent distribution is simply the average of the $H_i$, hence
the name `Average Shifted Histogram'.  Note that resulting estimate
(with $N$ times more bins than the original) is not a true histogram,
because the height of a `bin' is not necessarily equal to the number
of events falling in that bin.  However, it is a superior estimate of
the parent distribution, because the dependence of initial bin
position is essentially removed.  In the limit $N \rightarrow \infty$
the ASH is equivalent to placing a triangular shaped {\it kernel} of
probability about each data point $t_i$~\cite{Scott}.

\subsubsection{Fixed Kernel Estimation}

In the univariate case, the general kernel estimate of the parent distribution is given by
\begin{equation}\label{E:1dKEYS}
\hat{f}_0(x)=\frac{1}{n h}   {\sum_{i=1}^n}
K \left( \frac{x - t_{i}}{h} \right),
\end{equation}
where $\{t_i\}$ represents the data and $h$ is the smoothing parameter
(also called the {\it bandwidth}).  Immediately we can see that our
estimate $\hat{f}_0$ is bin-independent regardless of our choice of
$K$.  The role of $K$ is to spread out the contribution of each data
point in our estimate of the parent distribution.  An obvious and
natural choice of $K$ is a Gaussian with $\mu=0$ and $\sigma=1$:
\begin{equation}\label{E:kernel}
K(x) = \frac{1}{\sqrt{2 \pi}}e^{-x^2/2}.
\end{equation}
Though there are many choices of $K$, Gaussian kernels enjoy the
attributes of being positive definite, infinitely differentiable, and
defined on an infinite support.  For physicists this means that our
estimate $\hat{f}_0$ is smooth and well-behaved in the tails.

Now we concern ourselves with the choice of the bandwidth $h$.  In
Equation~\ref{E:1dKEYS}, the bandwidth is constant for all
$i$.  Thus, $\hat{f}_0$ is referred to as the {\it fixed kernel
  estimate}.  The role of $h$ is to set the scale for our kernels.
Because the kernel method is a non-parametric method, $h$ is
completely specified by our data set $\{t_i\}$.  In the limit of a
large amount ($n \rightarrow \infty$) of normally distributed
data~\cite{Scott}, the {\it mean integrated squared error} of
$\hat{f}_0$ is minimized when
\begin{equation}\label{E:hopt}
h^* = \left( \frac{4}{3} \right)^{1/5} \sigma n^{-1/5}.
\end{equation}
Of course, we rarely deal with normally distributed data, and,
unfortunately, the optimal bandwidth $h^*$ is not known in general.
In the case of highly bimodal data (\eg\ the output of a neural
network discriminate), the standard deviation of the data is not a
good measure for the scale of the true structure of the distribution.

\subsubsection{Adaptive Kernel Estimation}
An astute reader may object to the choice of $h^*$ given in
Equation~\ref{E:hopt} on the grounds of self-consistency -
non-parametric estimates should only depend on the data locally, and
$\sigma$ is a global quantity.  In order for the estimate to handle a
wide variety of distributions as well as depend on the data only
locally, we must introduce {\it adaptive kernel estimation}.  The only
difference in the adaptive kernel technique is that our bandwidth
parameter is no longer a global quantity.  We require a term that acts
as $\sigma_{\rm local}$ in Equation~\ref{E:hopt}.
Abramson~\cite{Abramson} proposed an adaptive bandwidth parameter
given by the expression
\begin{equation}\label{E:adaptive}
h_i = h/\sqrt{f(t_i)}.
\end{equation}
Equation~\ref{E:adaptive} reflects the fact that in regions of high
density we can accurately estimate the parent distribution with narrow
kernels, while in regions of low density we require wide kernels to
smooth out statistical fluctuations in our empirical probability
density function.  Technically we are left with two outstanding
issues: $i$) the expression for $h_i$ given in Equation~\ref{E:adaptive}
references the {\it a priori} density, which we do not know, and $ii$)
the optimal choice of $h$ has still not been specified.  Clearly, $h^*
\propto \sqrt{\sigma}$, because of dimensional analysis.  
Additionally, $\hat{f}_0$ is our best estimate of the true parent
distribution.  Thus we obtain
\begin{equation}\label{E:1dKEYSadaptive}
\hat{f}_1(x)=\frac{1}{n}   {\sum_{i=1}^n}
\frac{1}{h_i}K \left( \frac{x - t_{i}}{h_i} \right),
\end{equation}
with
\begin{equation}\label{E:hoptadaptive}
h^*_i =  \rho
\left( \frac{4}{3} \right)^{1/5} 
\sqrt{\frac{\sigma}{\hat{f}_0(t_i)}} n^{-1/5}.
\end{equation}

The adaptive kernel estimate can be thought of as a ``second
iteration'' of the general kernel estimation technique.  In practice,
the adaptive kernel technique almost completely removes any dependence
on the original choice of the bandwidth in the fixed kernel estimate
$\hat{f}_0$.  Furthermore, the adaptive kernel deals very well with
multi-modal distributions.  In extreme situations (\ie\ when the scale
of the local structure of the data $\sigma_{\rm local}$ is more than
about two orders of magnitude smaller than the standard deviation
$\sigma$ of the data) the factor $\rho$ in
Equation~\ref{E:hoptadaptive} should be adjusted from its typical
value of unity.  In that case
\begin{equation}\label{E:rho}
\rho = \sqrt{\frac{\sigma_{\rm local}}{\sigma}}.
\end{equation}
We have concluded the construction of a non-parametric estimate
$\hat{f}_1$ of an univariate parent distribution based on the
empirical probability density function.  Our estimate is
bin-independent, scale invariant, continuously differentiable,
positive definite, and everywhere defined.


\subsubsection{Boundary Kernels} \label{SS:UnivariateBoundaryKernel}
Both the fixed and adaptive kernel estimates assume that the domain of
the parent distribution is all of $\Bbb{R}$.
However, the output of a neural network discriminant, for example, is
usually bounded by $0<x<1$, where $f(x\le 0)=f(x\ge 1)\equiv 0$.  In
order to avoid probability from ``spilling out'' of the boundaries we
must introduce the notion of a {\it boundary kernel}.  Without
boundary kernels, our estimate will not be properly normalized and
underestimate the true parent distribution close to the boundaries.

Boundary kernels modify our traditional Gaussian kernels so that the
total probability in the allowed regions is unity.  Clearly, our
kernel should smoothly vary back to our original Gaussian kernels as we move
far from the boundaries.  This constraint quickly reduces the kinds of
boundary kernels we need consider.  Though a large amount of work has
been put forward to introduce kernels which preserve the criteria
\begin{equation}\label{E:criterea}
\int_{-\infty}^{\infty} t K(t) dt = 0,
\end{equation}
these methods do not suit themselves well for physics applications.
The primary problem is that the parametrized family of boundary kernels
may contain kernels that are not positive definite - which negates
their applicability to physics.  Also, boundary kernels satisfying
Equation~\ref{E:criterea} systematically underestimate the parent
distribution at a moderate distance from the boundary and overestimate
very near the boundary.
\begin{figure}\label{F:boundary}
\begin{center}
\epsfig{file=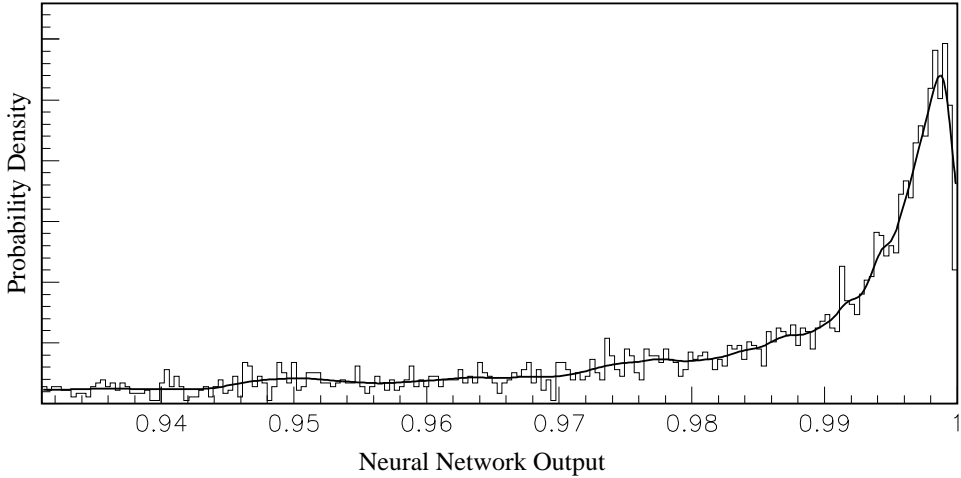, width=5 in}
\end{center}
\caption{The performance of boundary kernels on a Neural Network distribution with a hard boundary}
\end{figure}

An alternate solution to the boundary problem is to simply reflect the data
set about the boundary~\cite{Scott}.  In that case, the probability
that spills out of the boundary is exactly compensated by its mirror.

\subsection{Multivariate Kernel Estimation}

The general kernel estimation technique generalizes to {\it
  d}-dimensions~\cite{Scott}.  One choice for the {\it d}-dimensional
kernel is simply a product of univariate kernels with independent
smoothing parameters.  The following discussion will be restricted to the context of such product kernels.

\subsubsection{Covariance Issues}

When dealing with multivariate density estimation, the covariance
structure of the data becomes an issue.  Because the covariance
structure of the data may not match the diagonal covariance structure
of our kernels, we must apply a linear transformation which will
diagonalize the covariance matrix $\Sigma_{jk}$ of the data.  Ideally,
the transformation would remain a local object; however, in practice
such non-linear transformations may be very difficult to obtain.  In
the remainder of this paper, the transformation matrix will be referred
to as $A_{jk}$, and the $\{\vec{t}_i\}$ will be assumed to be
transformed.

\subsubsection{Fixed Kernel Estimation}

For product kernels, the fixed kernel estimate is given by
\begin{equation}
\hat{f}_0( \vec{x}  )=\frac{1}{n  h_1   \dots  h_d}   {\sum_{i=1}^n}  \left[
\prod_{j=1}^d K \left( \frac{x_j - t_{ij}}{h_j} \right) \right].
\end{equation}
In the asymptotic limit of normally distributed data, the {\it mean
  integrated squared error} of $\hat{f}_0$ is minimized when
\begin{equation}
h_j^* = \left( \frac{4}{d+2} \right)^{1/(d+4)} \sigma_j n^{-1/(d+4)}.
\end{equation}

\subsubsection{Adaptive Kernel Estimation}\label{SS:MultivariateAdaptiveKernel}

The adaptive kernel estimate $\hat{f}_1(\vec{x})$ is constructed in a
similar manner as the univariate case; however, the scaling law is
usually left in a general form.  Because most multivariate data
actually lies on a lower dimensional manifold embedded in the input
space, the effective dimensionality $d'$ must be found by maximizing
some measure of performance or making some assumption.  Thus the
multivariate adaptive bandwidth is usually written
\begin{equation}\label{E:ddhoptadaptive}
h_{i}=h f^{-1/d'}(\vec{t}_{i}).
\end{equation}
Though $d' \approx d$, the precise value depends on the problem.  Note
that the form of $h_i$ given in Equation~\ref{E:ddhoptadaptive} is
independent of $j$, thus it produces spherically symmetric kernels.
This is clearly not optimal.  Furthermore, when $d' \ne d$ the optimal
value of $h$ may vary wildly.  This is because the units are no longer
correct and $(d/d')$ powers of scale factors are introduced by
$f^{-1/d'}$.  Both of these problems may be remedied with the
introduction of a natural length scale associated with the data: the
geometric mean of the standard deviations of the transformed $\{t_i\}$,
$\sigma = \det(A_{}\Sigma_{}A^T_{})^{1/2d}$.  In the absence of
local covariance information, the best we can do is assume that the
$h_j$ are proportional to $\sigma_j$ and inversely proportional to
$f^{1/d'}$.  Thus we arrive at
\begin{equation}\label{E:h_ij}
h_{ij}^* =  \left( \frac{4}{d+2} \right)^{1/(d+4)} n^{-1/(d+4)} 
\left( \frac{\sigma_j}{\sigma}\right)
\sigma^{(1-d/d')}
f^{-1/d'}(\vec{t}_i),
\end{equation}
which is produces estimates that are invariant under
linear-transformation of the input space when the covariance matrix is
diagonalized.

\subsubsection{Multivariate Boundary Kernels}\label{SS:MultivariateBoundaryKernel}

Just as in the univariate case, it is possible that the physically
realizable domain of our parent distribution is not all of
$\Bbb{R}^d$, but instead a bounded subspace of $\Bbb{R}^d$.
Typically, this situation arises when one of the components of the
sample vector is bounded in the univariate sense (\ie~$t_j<x_j^{\rm
  max}$).  However, once we diagonalize the covariance matrix of our
data the boundary condition will take on a new form in the transformed
coordinates.  In general, any linear boundary in our original
coordinates $x_j$ can be expressed as $c_j x_j = C$, where $c_j$ is
the unit-normal to the ($d-1$)-dimensional hyperplane in our
$d$-dimensional domain and $C$ is the distance between the origin and
the point-of-closest approach.  After transforming to a set of
coordinates $x_j' = A_{jk} x_k$, in which the $\{\vec{t}_i\}$ have
diagonal covariance, our boundary condition is given by\newline 
$d_j x_j' = c_k A_{kj}^{-1} A_{jk} x_k = C$.
Thus, for each boundary one must introduce a reflected sample
$\{t_i^{\rm refl}\}$ with
\begin{equation}
t_{ij}^{\rm refl} = t_{ij} + 2(C-d_k t_{ik})d_j,
\end{equation} 
in order to rectify the probability that spilled into unphysical
regions.

\subsubsection{Event-by-Event Weighting}\label{SS:Event-by-Event}

In high-energy physics it is often necessary to combine data from
heterogeneous sources (\eg~independently produced Monte Carlo data
sets which together comprise the Standard Model expectation).  In
general one would like to estimate the parent distribution from a more
general {\it empirical probability density function} ${\rm epdf}(x) =
\sum_i w_i \delta(x-t_i)$, where $w_i$ represents the weight or {\it a
  posteriori} probability of the $i^{th}$ event.  In the case of combining
various Monte Carlo samples, one must reweight all events of a sample
to some common luminosity (say, 1 pb$^{-1}$) before combining them.  Thus for
a Monte Carlo sample with $n_{MC}$ events and cross-section $\sigma_{MC}$
each event must be weighted with $w_i = 1~{\rm (pb^{-1})} /\mathcal{L}_{\rm eff} = \sigma_{MC}
/ n_{MC}$



The covariance matrix of the weighted sample must be generalized as
follows:
\begin{equation}
{\Sigma_{jk}} = \sum_{i=1}^{n} \frac{(t_{ij} - \mu_j)(t_{ik} - \mu_k)}{n} 
\longrightarrow 
\tilde{\Sigma}_{jk} 
= \sum_{i=1}^{n} w_i \frac{(t_{ij} - \tilde{\mu}_j)(t_{ik} - \tilde{\mu}_k)}{\tilde{n}},
\end{equation}
where $\tilde{n} = \sum_{i=1}^{n} w_i$ and $\tilde{\mu} = \sum_{i=1}^{n}
w_i t_i / \tilde{n}$.  Then our estimate is simply given by

\begin{equation}
\hat{f}_1( \vec{x}  )=\frac{1}{\tilde{n}}   {\sum_{i=1}^n}
w_i
\left[ \prod_{j=1}^d \frac{1}{\tilde{h}_{ij}}  K \left( \frac{x_j - t_{ij}}{\tilde{h}_{ij}} \right) \right].
\end{equation}


\section{Applications}\label{S:Applications}

Kernel estimation techniques are applicable to all situations in which
it is necessary or useful to have an estimate of the parent
distribution of a set of data.  In high-energy physics the use of
kernel estimation runs the gamut from event selection to confidence level calculation.  

\subsection{Confidence Level Calculations}\label{SS:DiscriminantVars}

The most wide-spread use of kernel estimation techniques has been in
the context of confidence level calculations, primarily for the Higgs
Searches at LEP \cite{KEYS, L3keys, OPALkeys, DELPHIkeys,
  LEPHIGGSkeys}.  Each selected candidate event has associated with it
one or several discriminant variables (\eg~reconstructed Higgs mass,
neural network output, \etc).  Estimates of the distribution of
discriminant variables are constructed for the signal and background
processes, $\mathcal{L}_{\rm sig}$ and $\mathcal{L}_{\rm back}$
respectively.  These estimates are used to further discriminate
between candidate events via the log-Likelihood ratio
$\log(\mathcal{L}_{\rm sig}/\mathcal{L}_{\rm sig})$ or some other
`signal estimator'.  With the log-Likelihood in hand, confidence level
calculations transcend pure number counting and allow for a
more powerful statistical tool with which to interpret the data
\cite{CLFFT}.

\subsection{Measurements}

\subsubsection{Maximum Likelihood Fits}

Another context in which kernel estimation has been applied is the
measurement of physical constants via maximum likelihood fitting.
Traditionally, the log-Likelihood $\log \mathcal{L} = \sum_i \log
f(t_i; \vec{\alpha})$ is maximized with respect to the parameters
$\vec{\alpha} = (\alpha_1, \alpha_2, \alpha_3, \dots)$.  In this
context, $f(t_i; \vec{\alpha})$ is a parametrized model of the
physical situation.  In practice not all of the $\alpha_j$ are
`floated' or varied in the maximization routine, but instead many
parameters are `fixed' from some independent measurement.  While this
model incorporates empirical or theoretical information, it may make
unwanted assumptions about our data.  

For an example, let us consider the measurement of $\sin 2 \beta$ at a
$B$ factory.  The probability density of a CP decay recoiling from a
tagged $B$ ($\bar{B}$) meson is given by
\begin{equation}
f(t; \beta) = e^{-\Gamma |t|} (1 \pm \sin 2\beta \sin \Delta mt), 
\end{equation}
where $t$ is the time difference between the decay of the CP state and
the recoiling $B$ ($\bar{B}$) tagged meson with $\Delta z = \gamma
\beta c t$.  However, in an experiment we must take into account the
mistag rate $w$ and the resolution of $\Delta z$.  The standard
prescription is to measure $w$ and parametrize the resolution
distribution $R(\Delta z_{\rm true} - \Delta z_{\rm reco})$ with a
single (or double) Gaussian with bias~$\delta$ and variance~$\sigma$.
The final probability distribution is obtained via a convolution with
the resolution function and is of the form $f(t; w, \delta,
\sigma,\beta)= R(\delta, \sigma) \otimes f(t;w, \beta)$.  Now with $w,
\delta$, and $\sigma$ `fixed' we must `float' $\beta$ to make our
measurement~\cite{BaBarPhysicsBook}.  Here the form of $R$, while
justified, will have a systematic influence on the measured value of
$\sin 2 \beta$.  If, on the other hand, the resolution function $R$
was estimated via a non-parametric means (\ie~kernel estimation
techniques), then there would be no artificial influence on the
measurement and non-trivial resolution effects would be taken into
account automatically.

\subsubsection{Non-Parametric Parameter Estimation}

Knuteson proposed a non-parametric parameter estimation technique
based on kernel estimation of the joint density of feature vectors and
the parameters $\vec{\alpha}$ to be measured.  Via Bayesian statistics
the {\it posterior density} of the parameters given an observation is
obtained.  Then one estimates the parameters by simply maximizing the
{\it posterior} density~\cite{alphaPDE}.

\subsection{Discriminant Analysis}\label{SS:DiscriminantAnal}
The first uses of kernel estimation in high-energy physics were in the
context of new particle searches.  In order to create a discriminant
analysis, one constructs the discriminant $D(\vec{x})$ according to:
\begin{equation}\label{E:D}
D(\vec{x}) = \frac{\hat{f}_{\rm sig}(\vec{x})}{\hat{f}_{\rm sig}(\vec{x}) + \hat{f}_{\rm back}(\vec{x})}.
\end{equation}
It is easily seen that signal-like events get mapped to 1 and
background-like events are mapped to 0.  By placing a cut on
$D(\vec{x})$ a (possibly disconnected) region (bounded by the
corresponding $(d-1)$-dimensional contour) in the input space is
selected.  Because of the geometrical complexity of the selected
region, analyses of this type are more efficient than linear
discriminant analyses.  Furthermore, because of the intuitive nature
of Equation~\ref{E:D}, many prefer analyses of this type over
Artificial Neural Networks.  While these analyses may avoid the
black-box properties of Artificial Neural Networks, their usefulness is
limited to $d < 6$ due to the {\it curse of dimensionality}.  However,
with more intelligently chosen variables, the practical restriction of
$d < 6$ is not much of a limitation.  Discriminant analyses such as
these were met with much success in the search for the top quark at
D$\emptyset$~\cite{miettinen1, miettinen2, miettinen3}.

\subsection{Event Selection and Cut Optimization}

An obvious application of kernel estimation techniques is the
improvement and automation of cut optimization.  Given signal and
background samples, one typically varies cuts to find a region
$\mathcal{R}$ which maximizes some measure of performance (\eg~$S/B$,
$S/\sqrt{B}$, \etc).  Traditionally, the cuts are applied directly to
the sample events, thus only discrete values of the performance
measure are obtained.  This leaves the experimentalist in a bit of a
quandary because in the neighborhood of the optimal cut value the
performance measure tends to have many local extrema and is
constant between sample points.  Not surprisingly, the optimal cut
value is often just chosen by eye.  

The above method of cut optimization may be generalized by estimating
the performance not from the raw samples, but instead from the
estimates of the signal and background parent distributions,
$\hat{f}_{\rm sig}$ and $\hat{f}_{\rm back}$ respectively.  Then
instead of having discrete values of $S$ and $B$ we obtain continuous
functions $S(\mathcal{R})$ and $B(\mathcal{R})$.  Finally, we are left
with the more well-defined problem of finding the global maxima of a
continuously varying performance measure - in the case of $S/\sqrt{B}$
we obtain
\begin{equation}
  \mathcal{R} = {\rm argmax} \left \{  \frac{n_{\rm sig}\int_\mathcal{R}\hat{f}_{\rm sig}(x)dx}{\sqrt{n_{\rm back}\int_\mathcal{R}\hat{f}_{\rm back}(x)dx}} \right \}.
\end{equation}
In fact, it should be pointed out that with a finite sample size there
is an inherent uncertainty in $\hat{f}_{\rm sig}$ and $\hat{f}_{\rm
  back}$.  Thus, there is an an inherent uncertainty in~$\mathcal{R}$.
In that case, the optimal selection strategy becomes probabilistic -
events on the boundary $\mathcal{R}$ should be selected with some well
defined probability.  While this strategy can be carried out in a
deterministic fashion (\eg~generating pseudo-random numbers with event
and run number as a random seed) the gain in performance does not
merit the added complexity of the selection.

\section{Available Packages}
\subsection{Univariate Packages}
\subsubsection{KEYS}
In order to implement univariate kernel estimation techniques to
produce FORTRAN functions of $\hat{f}_1$ based on the data in HBOOK
ntuples, {\it KEYS} was developed.  {\it KEYS}
\footnote{{\it KEYS} is an acronym for Kernel Estimating Your Shapes} 
has been implemented in a Perl script which: 1) parses an
input file specifying the input ntuples and output filenames of a set
of shapes,
\footnote{The word {\it shape} is lingo for the continuous
 estimate of the parent distribution}
 2) interfaces with PAW to
obtain the data set $\{t_i\}$, 3) produces a FORTRAN function to
calculate $\hat{f}_1(x)$, 4) compiles and evaluates the FORTRAN
function, 5) produces plots (see Figure~2) 
of the estimate and the Kolmogorov-Smirnov probability that the
$\{t_i\}$ is compatible with $\hat{f}_1$~\cite{KS}, and 6) produces a second
FORTRAN function which linearly interpolates $\hat{f}_1$ between 300
sample points %
in order to expedite the evaluation of the function.  {\it KEYS} has
been used extensively for the parametrization of discriminant variable
distributions in Higgs searches at LEP \cite{KEYS, ALEPHkeys, OPALkeys,
  DELPHIkeys, L3keys, LEPkeys}.

\begin{figure}\label{Fig:KEYSoutput}
\begin{center}
\epsfig{file=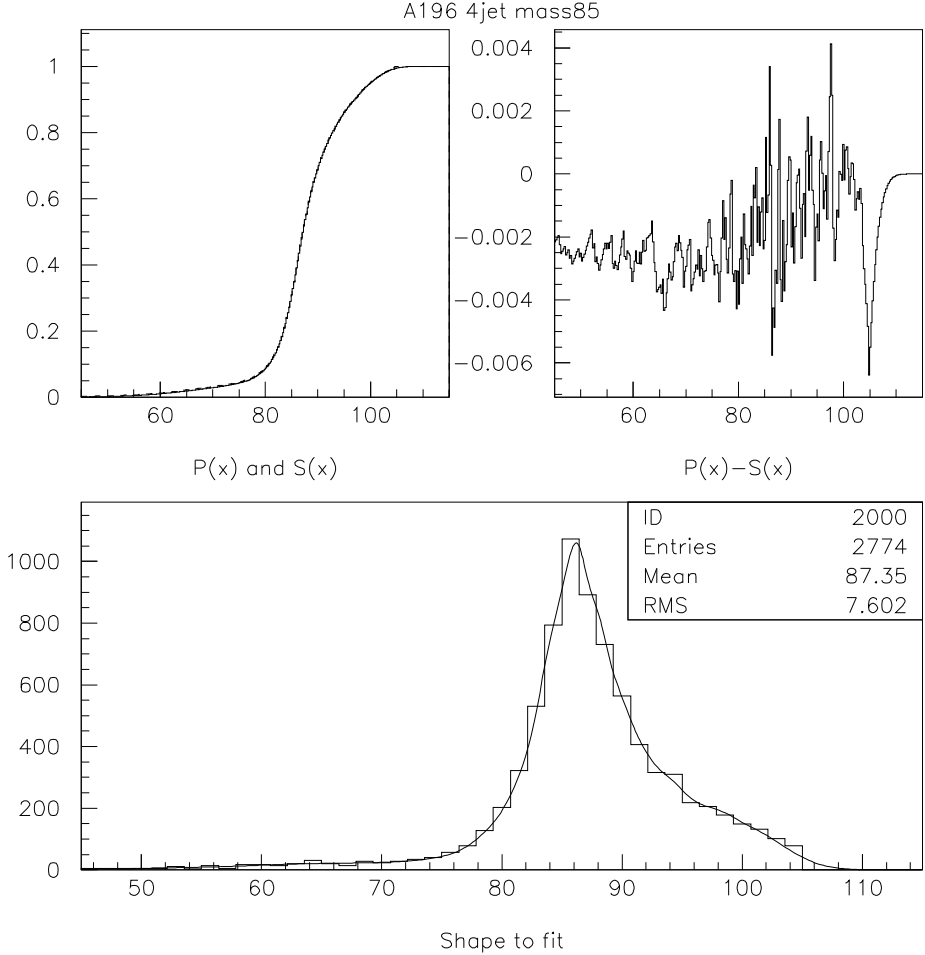,  height=5 in}
\caption{The standard output of the {\it KEYS} script.  The top left plot shows the cumulative distributions of the {\it KEYS} shape and the data.  The top right plot shows the difference between the two cumulative distributions, the maximum of which is used in the calculation of the Kolmogorov-Smirnov test.  The bottom plot shows the shape produced by {\it KEYS} overlayed on a histogram of the original data.}
\end{center}
\end{figure}


\subsubsection{HEPUKeys}

Another implementation of univariate kernel estimation has been
developed by Jeremiah Mans of the L3 collaboration.%
\footnote{electronic mail: Jeremy.Mans@cern.ch}  %
This implementation is written in \cpp.  The {\it HEPUKeys} class
executes a C-compiler, and uses the dynamic loading methods (dlopen,
dlclose, dlsym) for interactive or embedded implementations of the
kernel estimation technique.  This embedded structure allows one to
include confidence level calculation, cut optimization, or any other
application of kernel estimation directly in the analysis code.

\subsection{Multivariate Packages}

\subsubsection{PDE}

The original implementation of multivariate kernel estimation in
high-energy physics was developed at Rice University by Holmstr\"{o}m, 
Miettinen%
\footnote{electronic mail: miettinen@physics.rice.edu} %
and Sain with guidance from Scott around 1994.  This VMS-based FORTRAN
implementation, called {\it PDE}, was applied to searches for the top
quark at D$\emptyset$ \cite{miettinen1, miettinen2, miettinen3}.

\subsubsection{RootPDE}\label{SS:RootPDE}

In order to produce a more portable and flexible package for kernel
estimation, the author developed%
\footnote{This work was partially funded by an {\it Envision} grant
  from Rice University}%
an object oriented implementation written in \cpp. The core numerical
procedures were left in a very general form with polymorphic kernels
and minimal user interface.  The core procedures have been adopted
by two complete packages: {\it RootPDE} and {\it WinPDE}.

Padley and Askew%
\footnote{electronic mail: padley@physics.rice.edu, askew@physics.rice.edu} %
interfaced the core numerical procedures described above into a shared
Root library called {\it RootPDE}.  This package was primarily
intended for discriminant analyses as described in Section
\ref{SS:DiscriminantAnal}.  However, this package was submitted to
D$\emptyset$'s CVS repository prior to the development of Equation~\ref{E:h_ij} (see Section \ref{SS:MultivariateAdaptiveKernel}) and
does not support event-by-event weighting (see Section
\ref{SS:Event-by-Event}) or boundary kernels (see Sections
\ref{SS:UnivariateBoundaryKernel} and
\ref{SS:MultivariateBoundaryKernel}).  

Since the original release of {\it RootPDE}, the author has been
including more advanced features and developed methods which allow a
kernel estimate object to write out a \cpp~function for use in other
applications.  Furthermore, the interface has been generalized from
pure discriminant analysis to a form more amenable to any of the applications described in Section \ref{S:Applications}.  Currently, this development is
taking place at \BaBar; however, it will be made publicly available
in the near future.

\subsubsection{WinPDE}

{\it WinPDE} is very similar to the original release of {\it RootPDE}
described in Section \ref{SS:RootPDE}; however, the interface is
written in VisualBasic and intended for Microsoft Windows platforms.
Furthermore, {\it WinPDE} can interface with Microsoft Excel
spreadsheets and offers a very intuitive user-interface.

\section{Conclusions}

\subsection{Comparison with SMOOTH}

It seems appropriate to put kernel estimation techniques in a proper
setting before concluding with a discussion of their inherent benefits.
While kernel estimation techniques may be applied to situations in
which a parametric estimates are popular, that comparison has
essentially been made by Section~\ref{SS:Overview}.  Instead, let us
consider perhaps the most widely used non-parametric density
estimation technique in high-energy physics: PAW's SMOOTH utility.


\subsubsection{SMOOTH}
A full development of the HQUADF function that is used by PAW's SMOOTH
utility is beyond the scope of this paper.  However, a brief outline
of the algorithm is presented.  First and foremost, it is important to
realize that SMOOTH operates on histograms and not on the original
data set $\{t_i\}$.  Thus, SMOOTH is dependent on the original binning
of the data.  SMOOTH was introduced by this journal in John Allison's
1993 paper 
\cite{Allison}.  We will restrict ourselves to the univariate case.
Essentially SMOOTH works by finding the bins $l$ of {\it significant
  variation} in the histogram $\{h_l\}$ and then using those points to
construct a smoothed linear interpolation.  Bins of significant
variation are those which satisfy $S_l > S^*$, where $S^*$ is a
user-defined significance threshold and
\begin{equation}
S_l = \left| \frac{h_{l+1}-2h_l+h_{l-1}}
{\sqrt{Var(h_{l+1})+4Var(h_l)+Var(h_{l-1})}}\right|.
\end{equation}
With the points of significant variation $\{x_l\}$ in hand, the
smoothed shape is given by
\begin{equation}
s(x) = \sum_l a_l \phi_l (|x-x_l|),
\end{equation}
where $\phi_l(r) = \sqrt{r^2+\Delta_l^2}$ are the radial basis
functions.  The $\Delta_l$ are user-defined smoothness parameters
(radii of curvature).  The $a_l$ are found by minimizing the $\chi^2$
between $s(x)$ and the original histogram.  As Allison pointed out
``lower $\chi^2$ can be obtained by reducing the cut on $S_l$ at the
expense of following more of what {\it might} only be statistical
fluctuations.''  By a different choice of $S^*$ and $\Delta$, the user
has the power to magnify or remove statistical fluctuation in the
data.

\subsubsection{Comparison}
Despite the user-specified parameters $S^*$ and $\Delta$, SMOOTH is a
non-parametric estimate of a probability density function based on a
set of data.  The primary differences between SMOOTH and kernel
estimates are their approach and their rigor.  While kernel estimates
are bin-independent {\it constructions} of the estimate, SMOOTH is a
parameter-dependent {\it fit} of the estimate to a user-provided
histogram.  Practically speaking, kernel estimates are based on well
defined statistical techniques and SMOOTH's estimates are adjusted by
eye allowing for user bias and large systematic uncertainty.


\subsection{Systematic Errors}

When kernel estimation techniques are applied to confidence level
calculations or parameter estimation, systematic effects become of
particular importance.  One may loosely classify the systematic errors
associated with probability density estimation as either inherent or
user-related errors.  In its pure form kernel estimation techniques
are entirely deterministic and have no user-specified parameters.  If
one decides to free the value of $\rho$ from its nominal value of
unity (see Equation~\ref{E:rho}) or allow $d' \ne d$, then
user-related systematic error are introduced.  For SMOOTH, the
user-related parameters $S^*$ and $\Delta$ can not be avoided.  In
addition to the possible user-related systematic errors, there are
inherent systematic errors introduced by any probability density
estimation technique.  For parametric estimates, this inherent
systematic is related to the quality of the model; while for
non-parametric estimates, this inherent systematic is related to the
flexibility of the technique.  The development of kernel estimation
techniques has been directly focused on flexibility and the
minimization of a particular choice of inherent systematic error: the
asymptotic mean integrated squared error \cite{Scott}.

In practice, an experimentalist will want to choose their own estimate
of the inherent systematic error (\eg~the effect on the measured value
of a parameter or 95\% confidence level limit).  This can be done in a
variety of ways that effectively reduce to producing a family of
estimates from independent samples of the same parent distribution.
This family may be obtained by simply splitting up the data or via toy
Monte Carlo simulation.  Because the systematic error introduced by
the estimation technique is a function(al) of the sampled parent
distribution (which is unknown), the estimate itself is the best
available choice of the parent distribution to be sampled in a Monte
Carlo study.

\subsection{Remarks}

Obviously, kernel estimation techniques are very powerful and very
relevant to high-energy physics.  While these techniques have been
applied to a wide range of analyses, they seem to by largely unknown
by the community.  It is the aim of this paper to describe kernel
estimation techniques, their applications, and their current
implementations in order to promote their use.  It is to the author's
personal gratification to be part of a cross-polination of ideas
between fields - in this case statistics and high-energy physics.

\section{Acknowledgements}
The author would like to thank Hannu Miettinen - without his original
guidance this paper would not have been written.  The development
of {\it KEYS} was largely influenced by members of the LEPHiggs
Analysis Working Group, most notably Chris Tully, Steve Armstrong,
Peter McNamara, Jason Nielsen, Arnulf Quadt, Tom Junk, Peter
Igo-Kemenes, Tom Greening, and Yuanning Gao.  Thanks to Paul Padley
and Andrew Askew for the development of {\it RootPDE} and Stephen
McCaul for aid in the design of its core numerical procedures.  Karl
Berkelman provided very useful discussion on the sensitive issues of
systematic errors.  Finally, the author is grateful for the support of
Sau Lan Wu during a stay at CERN in which the majority of this work
was conducted.

\end{document}